%Paper: hep-th/9208018
%From: STEIF@binah.cc.brandeis.edu
%Date: Wed, 5 Aug 1992 17:59 EDT
%Date (revised): Thu, 27 Aug 1992 16:40 EDT

\input harvmac

\Title{BRX-TH-335}
{{\vbox {\centerline{
Gravity Theories with   Lightlike Sources in $D=3$
}}}}

\bigskip
\centerline{S. Deser}
\centerline{Alan R. Steif}
\bigskip\centerline
{\it Brandeis University}
\centerline{\it Department of Physics}
\centerline{\it Waltham, MA  02254}
\centerline{\it deser@binah.cc.brandeis.edu}
\centerline{\it steif@binah.cc.brandeis.edu}
\vskip .2in

\noindent
ABSTRACT: Spacetimes generated by a lightlike particle source
for  topologically massive gravity
and its limits --- Einstein gravity
and the pure gravitational Chern-Simons model --- are obtained both
by solving the field
equations and by infinite boosts of   static metrics.  The resulting
geometries are the first known  solutions of topologically massive gravity
 that are asymptotically flat and generated
by compact matter sources.  Explicit metrics describing various
multiphoton solutions are also derived. For Einstein gravity, we also construct
such solutions by null boost identifications of Minkowski space and thereby
obtain limits on the energies of the sources.

\Date{}

\def\a{\alpha}
\def\b{\beta}
\def\d{\delta} 
\def\e{\epsilon}

\def\k{\kappa}
 \def\L{\Lambda}
\def\m{\mu}
\def\n{\nu}

\def\p{\pi} 
\def\r{\rho}

 \def\O{\Omega}

\def\pa{\partial}
\def\ra{\rightarrow}
\def\pr{\prime}
\newsec{Introduction.}

There has been renewed interest in geometries generated by a lightlike particle
in general
relativity because of their relevance
to quantum scattering of two gravitationally coupled particles
 at   Planckian energies {\ref\thooft{ T. Dray and G.
't Hooft, {\it Nucl. Phys.} {\bf B253} (1985) 173; G. 't Hooft, {\it Phys.
Lett.} {\bf B198} (1987) 61;  {\it Nucl. Phys.} {\bf B304}
(1988) 867.
}}.
Three-dimensional gravity theories provide a simplified, but interesting,
arena
to study these questions: in Einstein gravity and (conformally invariant)
Chern-Simons gravity (CSG)  there are no field degrees
of freedom at all, while the more general topologically massive gravity (TMG)
{\ref\tmgt{ S. Deser, R. Jackiw, S. Templeton, {\it Ann.  Phys.} {\bf
140} (1982) 372.}}
is dynamical, yet very different from D=4 gravity.
We propose
here to solve for and interpret the  geometries generated by lightlike
matter sources (which we call ``photons" for short), as a first step
towards studying the corresponding quantum scattering problems in
these systems {\ref\dms{S. Deser, J. McCarthy, A. Steif, ``Planck Scale
Scattering in $2+1$ Gravity Theories'', in preparation. }.
In general, photons are described by plane-fronted wave spacetimes. We
will exploit the fact that  such
metrics ``linearize" the field equations (so
that the linear and full curvature tensors coincide) to obtain a wide class of
solutions.
In particular, we
will find, for full nonlinear TMG, the metric generated by a photon; this
is the first known example of an asymptotically flat
solution with a localized source in the theory.
Indeed, in TMG only the linearized
``Schwarzschild" solution is known for a static particle;
in the limit of infinite boost it
will be shown to coincide with the exact plane wave metric. We will also
display
multiphoton solutions in these models, and relate the geometrical aspects of
our results to those discussed in
{\ref\djt{S. Deser, R. Jackiw, G. 't Hooft, {\it Ann. Phys.} {\bf 152}
(1984)  220.}} for pure Einstein gravity.

{\newsec{The Single Photon Solution.}}

The plane-fronted wave ansatz in $D$ spacetime dimensions,
\eqn\plane {ds^2      = ds^2_0 + F(u,\vec{y})du^2,
\;\;\;ds^2_0  \equiv   -dudv + d\vec{y}^2, }
 simplifies the
Einstein tensor to the linearized form
\eqn\Ein {G_{\mu\nu} =-{1\over 2} \nabla^2_T F~~ l_\m l_\n \;\; , \;\;\;
l_\m \equiv  \partial_\m \, u . }
Our sign convention is $R_{\m\n}\sim
 + \partial_{\a}\Gamma^{\a}_{\m\n},$
$\vec{y}$ are the $D-2$ transverse coordinates,
$u= t-x$ and $v=t+x$ are the usual
lightcone coordinates, and
$\nabla^2_T$ is the flat space Laplacian in the
transverse dimensions. Henceforth, we specialize to $D=3 $ and denote
$y$-differentiation by a prime.   Thus, only the component
$G_{uu}  =  -{1\over 2}F^{\prime\prime}$, or
equivalently $R_{uyuy}$ (since the Riemann and  Einstein tensors
are double duals of one another)
 fails to vanish. We shall also need  the Cotton (Weyl) tensor,
$C^{\mu\nu} \equiv  \e^{\m\a\b} D_\a (R_\b~^\n -
\textstyle{1\over 4} \d^\n_\b R)$; it  too simplifies drastically:
\eqn\cotton{C_{\mu\nu} = - {1\over 2} F^{\prime\prime\prime} l_\m l_\n }
where our volume form convention is $\e^{txy}= 1. $
The (parity violating) field equations of TMG {\tmgt} are
\eqn\eom{E_{\mu\nu} \equiv G_{\mu\nu} + {1\over \m} C_{\mu\nu} =
-\kappa^2 T_{\mu\nu}}
where $\k^2$ is the Einstein
constant with dimensions of inverse mass and $\m $ is the
graviton's mass or, in our classical framework,
its inverse range. For simplicity of notation we take $\m >0$; the
theory with opposite sign
choice is parity conjugate to {\eom} and goes through analogously.
Note that TMG reduces to Einstein gravity\footnote{*}
{\rm More precisely, it reduces to ``ghost'' Einstein gravity, with the
opposite sign of $\k^2$ than that required in $D=4$; this sign choice is forced
by requiring TMG to be non-ghost {\tmgt}.}   as $\m\ra\infty$, and to CSG as
$\m\ra 0$, $\k^2\ra\infty$ with $\k^2\m$ fixed.

For our plane wave ansatz and  with a right-moving photon source of energy
$E$, {\eom}
reduces to  just its $uu$ component,
\eqn\eomF{( 1+ {1\over \m} {\partial\over \partial y} )F^{\prime\prime}
=  2 \k^2 E\d ({y}) \d (u) \; .}
Hence the curvature is
\eqn\riemann{G_{uu} = R_{uyuy} = -{1\over 2} F^{\prime\prime}=
 - \m\k^2  E e^{-\m y} \theta(y)\delta(u) + A(u)e^{-\m y}.}
We will drop the homogeneous solution, $Ae^{-\m y}$, to maintain asymptotic
flatness.  From {\riemann} we see
that the spacetime is flat off the $u=0$ plane and for
negative $y$.
Integrating {\riemann} yields  the metric\footnote{**}
{\rm This method of solving {\eom} was found independently by J. McCarthy; see
{\dms}.}
\eqn\eff{F= 2\k^2 Ef(y)\delta(u)  + By + C , \;\;\;
 f= (y+ {1\over \m}(e^{-\m y}-1))\theta (y).}
Here $f$ can be viewed as an inverse of the operator
$(1+ {1\over\m}{\partial\over\partial
y})({\partial\over\partial y})^2$; the homogeneous solutions involve
arbitrary functions $B(u),C(u)$
that can be removed by the coordinate transformation
\eqn\trans{\eqalign{u & \rightarrow u\cr
                    v  & \rightarrow v+  by -{1\over 4} \int_0^u b^2 +  {1\over
2} b\int_0^u b + c \cr
                    y  & \rightarrow y + {1\over 2} \int_0^u b \cr}}
where $b(u) = \int_0^u B$ and $c(u) = \int_0^u C.$
 If we make a
further coordinate transformation to remove the $\d (u)$ factor in $F$,
$\tilde v  = v - 2\k^2 E f(y) \theta (u),$
  then the metric  {\plane } with {\eff} takes the  form
\eqn\gravid{\eqalign {ds^2  &= -du d\tilde v + dy^2
 -2\k^2 E  \theta(u) f^{\prime}(y) dudy\cr
&= \theta (u)  \{ -du d (\tilde v + 2\k^2 E f(y))  + dy^2  \}
 +  \theta (-u)  \{ -dud\tilde v + dy^2 \} .\cr  }  }
This geometry clearly  corresponds
 to taking the two flat halfspaces, $u>0$ and $u<0$ in $R^3$
and identifying the points
$(u,\tilde v,y) = (0,\tilde v_0,y_0)$ on $u=0^-$ with
the points $(u,\tilde v,y) =
(0,\tilde v_0 + 2 \k^2 E f(y_0) ,y_0)$ on $u=0^+$.

{}From the solution to the TMG model, {\eff},  we can of course recover its two
limits ---  (ghost) Einstein gravity and CSG --- or  more simply,
 we can obtain them directly from
{\eomF}. In the case of  Einstein gravity, we are more interested in the
usual non-ghost theory to compare with previous work; hence, {\it in this limit
only,} we solve the equations $G_{\m\n} = + \k^2 T_{\m\n}$ corresponding to
$\m\ra\infty$ with $\k^2\ra - \k^2$:
\eqn\puregravityriemann{G_{uu} = R_{uyuy} = -{1\over 2} F^{\prime\prime}= \k^2
 T_{uu} =
 \k^2  E  \d (y)\delta(u)  ,}
vanishes  except on
the photon trajectory. Integrating {\puregravityriemann} we find
\eqn\puregravityeff{F= -2 \k^2 E y\theta (y)\d (u) + B(u)y + C(u),}
corresponding to the $\m\ra \infty$ $(\k^2\ra -\k^2)$ limit of {\eff}.
(This solution with $(B,C)= (\k^2E\d (u),0)$ has appeared previously
{\ref\ccv{A. Cappelli, M. Ciafaloni, and P. Valtancoli, {\it Phys. Lett.} {\bf
B273} (1991) 431.}}.)
The homogeneous $By+C$ terms can again  be removed by {\trans},
 leaving the interval
\eqn\puregravitymetric { ds^2 = ds_0^2 -2  \k^2 E y\theta (y) \d (u) du^2 }
where $F$ now has support on the $\{ u=0,\, y>0\}$ null halfplane.
In contrast with TMG  $F$ can also  be made to
have support just on the $y<0$ halfplane by choosing
$(B,C) = ( 2\k^2 E \d (u), 0)$. Both forms of the metric are of course
flat outside the source and  related by the coordinate transformation,
{\trans}.

The CSG model is likewise most easily solved directly from {\eomF}, i.e., from
$F^{\prime\prime\prime} = 2\bar\m E \d(y) \d (u)$ where
$\bar\m \equiv \k^2 \m$ is the relevant coupling constant.
   Furthermore, since $C^{\m\n}$ is the $D=3$ Weyl tensor,
it only determines the metric up to a conformal factor; thus using {\eomF} or
the corresponding limit of the TMG solution involves a particular conformal
gauge choice.  Here the curvature is
\eqn\csgriemann{G_{uu} =  -{1\over 2} F^{\prime\prime}=
-\bar \m  E ( \theta (y) \delta(u)  +  D(u) ) .}
There is no preferred choice of $D(u)$, since the curvature cannot
 in any case  be made
to vanish at both  $y=\infty$ and $y=-\infty$. The corresponding metric
(after removing the homogeneous solutions  $By+C$ and up to an arbitrary
conformal factor) is
\eqn\csgmetric{ds^2 = ds_0^2 +\bar \m E y^2(\theta (y)\d(u) + D(u))du^2.}

{\newsec{Boosting.}}

In this section, we obtain the one-photon solutions
by an infinite boost of the
corresponding static  ``Schwarzschild'' metrics in the same
spirit as for $D=4$ gravity
{\ref\as{P. C. Aichelburg and R. U. Sexl,
{\it J. Gen. Rel. Grav.} {\bf 2} (1971) 303.}}.
We begin with the usual non-ghost
Einstein gravity, whose conical metric for a static mass $m$ {\djt} we
write in the form:
\eqn\cone{ds^2=ds_0^2 +  (\a^2-1) {(ydx-xdy)^2\over (x^2 +y^2)},}
where $ \a= 1- {\k^2 m\over 2\p}$, and $ds_0^2 =  -dt^2  + dx^2 + dy^2.$
We now perform a boost in the $x$ direction with boost parameter $\b = {\rm
tanh}\;\xi.$ The first term, $ds_0^2$, is of course invariant,
while in the second term $x$ is replaced by
$\tilde x \equiv {\rm cosh}\;\xi\, x - {\rm sinh}\;\xi\, t .$
For large $\xi$ and small $m$ with the energy $E=m\,
{\rm cosh}\;\xi$ fixed,
the metric becomes
\eqn\largeboost{ ds^2 = ds_0^2 - {\k^2 E }  {e^{\xi}\over 2\p}
 { (-ydu +udy)^2\over
 ({e^{\xi}\over 2} u)^2  +  y^2 } .}
Using the fact that
\eqn\integrala{ \lim_{\xi\ra\infty} {e^{\xi}\over 2\pi }
 {1\over ({e^{\xi}\over 2} u)^2 + y^2}    ={1\over \pi}
 \d(u)  \int_{-\infty}^{\infty} {ds\over  s^2+y^2} = {1\over |y|} \d (u), }
 we see that  {\largeboost} reduces to   {\puregravityeff}, the one-photon
solution, with
$(B,C)= (\k^2 E\d (u),0) $. It is not surprising that we
obtained  the form of the metric
 $(F\propto |y|\d(u))$ symmetric in $y$, since
we began with a symmetric metric {\cone} and boosted in $x$.

We now obtain the one-photon TMG metric by boosting the corresponding
stationary
``Schwarzschild'' TMG solution.
Actually only the linearized metric for  a static source is known
{\ref\anyon{S. Deser, {\it Phys. Rev. Lett.} {\bf 64} (1990) 611.}}.
This will be sufficient, however, since we take $m\rightarrow 0$ in
the infinite boost method.
Indeed, the linearized solution is always a sufficient starting point
(for Einstein gravity recall the existence of
Eddington coordinates in terms of which the Schwarzschild solution
is already linear in $m$).
It is more convenient here to first
apply the coordinate transformation $\rho = {r^{1+ { \k^2 m\over
2\pi}}\over {1+ {\k^2 m\over 2\pi}}} \approx r
+ {\k^2 m\over 2\p} r ({\rm ln}\;r -1)$ to
the linearized solution as given in {\anyon}.
 We then have
\eqn\metric{  ds^2  =  ds_0^2   + {1\over 2\pi} \k^2 m\{ K_0(\m\rho)
( dt^2 + d\rho^2 +\rho^2d\phi^2) + 2\rho^2 d\phi^2  + {2\over \m}
(\rho{\partial K_0(\m \r)\over \partial\rho} +1 )dtd\phi \} }
 where $ K_0 $
is the modified Bessel function and $ds_0^2 = -dt^2 +d\r^2 + \r^2 d\phi^2 .$
[In the limit $\m\ra \infty$, we of course regain the (linearized)
``ghost'' version of
{\cone} in  polar coordinates.]
For a boost with large $\xi$
\eqn\glim{\eqalign{ t,x &\ra {e^{\xi}\over 2} u,\;\;\;
\r^2\ra \tilde\r^2  \equiv
 ({e^{\xi}\over 2}  )^2u^2 + y^2  \cr
  dt^2 \ra ({e^{\xi}\over 2} )^2du^2,\;\;\r^2 d\phi^2 &\ra
 ({e^{\xi}\over
2})^2 { ( udy  - ydu )^2\over \tilde\r^2  } ,\;\; dtd\phi\ra
 ({e^{\xi}\over
2})^2 { ( udy -ydu)du \over \tilde\r^2  }
.\cr}}
{}For large $\xi$ and small $m$ with $E= m\, {\rm cosh}\;\xi$ fixed,
eqn.  {\metric}  then becomes
\eqn\boostedmet{  ds^2  =  ds_0^2   +
{1\over \pi} \k^2 E ({e^{\xi}\over 2}) \{  K_0(\mu\tilde\rho) du^2
 + {(udy -ydu)^2 \over \tilde\rho^2}   + {1\over \mu}
(\tilde\rho{\partial K_0(\m\tilde \r)\over \partial\tilde\rho} +1 ){(udy -ydu)
du \over
\tilde\r^2 }  \}  .}
To evaluate this limit, we recall the identity generalizing {\integrala},
\eqn\generallimit{  \lim_{\xi\ra\infty} {e^\xi\over 2} W
( \tilde\r^2   ,y)  = \d (u)
 \int_{-\infty}^\infty W (s^2 +y^2,y)\,\, ds,\;\;\; \tilde\r^2
\equiv   ({e^{\xi}\over 2}  )^2u^2 + y^2 }
provided $\tilde \r W (\tilde\r^2 ,y)\ra 0$ as $\tilde \r\ra\infty .$
It follows from the $\d$-function in {\generallimit} that only the coefficient
of  $du^2$ in {\boostedmet} survives the limit.
Upon performing the required integration in {\generallimit},
 we check that the infinite boost limit of {\boostedmet}
is just  the one-photon TMG solution,  {\eff}, with
$B= -\k^2 E\d (u),$ $C= {\k^2 E\over \m}\d (u) ;$
again $B$ and $C$ can be transformed away by {\trans}.

Finally, we remark that there is
no direct boost approach to CSG since $C_{\m}^{\m}$
vanishes identically,  and hence $C^{\m\n}$
can only couple to a traceless stress tensor,
such as a lightlike particle; TMG can thus be thought of as a
``regularization'' permitting us also
to obtain the CSG metric by first boosting and then limiting to CSG.

{\newsec{Multiphoton Solutions.}}

We  now derive multiphoton  solutions to TMG, as well as to
Einstein gravity.
Surprisingly, we will find that there exist geometries, even in the dynamical
context of
TMG, corresponding to some
superpositions of freely moving, noncollinear photons. We therefore begin by
describing the kinematics of
$n$ photons with energies $E_i, i=1,\dots, n$ moving along
lightlike
flat space geodesics, {\it i.e.,} straight lines in Cartesian coordinates with
null tangent vectors. Every photon's worldline  can be obtained by a rotation
and translation, $x^{\m} \ra x^{\m}_i = (\O_i x)^{\m} + a_{i}^{\m}$,
 of some base trajectory $u\equiv t-x=0,\, y=0.$
 The corresponding energy-momentum
tensor is $T_{i\m\n} = E_i\d (u_i) \d
(y_i) \pa_{\m}u_i \pa_{\n}u_i.$
The solution to the associated, one-photon, equation of motion
$E_{\m\n}  = -\k^2  T_{i\m\n}$  is correspondingly
obtained by applying the rotation and translation
to the base source  one-photon solution of Section 2, yielding
 $g_{i\m\n}   =\eta_{\m\n}  + 2\k^2 E_i f(y_i) \d (u_i) \pa_{\m}u_i
\pa_{\n}u_i.$
To understand the $n$-photon system, first consider two photons
whose energy-momentum tensor is given by the sum of the individual one-photon
contributions,  $T_{\m\n} = T_{1\m\n} +T_{2\m\n}.$ For parallel photons, which
we can assume to be moving along the $x$-axis, each of their
 one-photon solutions,
$g_{1\m\n}$ and $g_{2\m\n}$, are of the form {\plane}. In this case
$E_{\m\n}$
linearizes\footnote{*}
{\rm We work with $h_{i\m\n} \equiv  g_{i\m\n} -\eta_{\m\n} $ rather than
$g_{i\m\n}$ in order to avoid ``overcounting'' their $\eta_{\m\n}$ parts.}
  $(E_{\m\n} [ h_{1\a\b}+ h_{2\a\b}] = E_{\m\n} [ h_{1\a\b}] +
E_{\m\n} [ h_{2\a\b}] )$, and the two-photon solution
is simply the superposition
of the one-photon solutions, $h_{\m\n} = h_{1\m\n} + h_{2\m\n}$. Superposition
of parallel photons also holds in $D=4$ gravity. However, in $D=3$
the one-photon solutions will be seen to
superpose for some non-parallel photon configurations  as well.
 The local functional $E_{\m\n} $ obeys
$E_{\m\n}[ h_{\a\b}=0] =0$.
Therefore, if the two regions on which $h_{1\a\b}$ and $h_{2\a\b}$ have
support are {\it disjoint, }
then $ E_{\m\n} [ h_{1\a\b}+ h_{2\a\b}] = E_{\m\n} [ h_{1\a\b}] +
E_{\m\n} [ h_{2\a\b}] $ implying again that the two-photon solution is given
  by $h_{\m\n} = h_{1\m\n} + h_{2\m\n}.$
[In $D=4$, the one-photon metric {\as} has support on an entire null
hyperplane; since two non-parallel hyperplanes necessarily intersect,
only parallel photons can be superposed there.] Now for general $n$,
superposing the
one-photon solutions, $g_{\m\n} = \eta_{\m\n} + \sum_{i=1}^n h_{i\m\n},$
will yield the $n$-photon
solution provided every pair of photons is either parallel or oriented such
that the respective halfplanes ($u_i =0,\,y_i>0$) on which the $h_{i\m\n}$ have
support  are disjoint.

For two non-colliding photons in Einstein gravity,
 this method of superposition can be used to
construct  the solution for
every orientation of the particles. For two photons in TMG as well as for $n>2$
photons in any of the models, not all orientations of the photons satisfy the
criteria for superposition;  in fact, no orientation of $n>4$ mutually
nonparallel photons satisfies them in TMG {\it or} in
Einstein gravity (as is geometrically clear).
Consider first two photons in Einstein gravity. We can go to their
center-of-momentum frame where we take the two particles to be
moving antiparallel in
the $x$-direction with equal energies $E$ along the trajectories  $u=y=0$  and
$v=0,\,y=a\neq 0.$
In the notation of the previous paragraph,
$u_1=u,\, v_1=v,\,y_1=y$ and $u_2=v,\, v_2=u,\,y_2= a-y.$
As discussed in Section 2, we can construct the Einstein gravity
solution for one photon,
$h_{\m\n}$,  moving along the $u=y=0$ trajectory to have support on either the
$\{ u=0, y>0 \}$ or $\{ u=0, y<0 \}$ null halfplanes by appropriate choice of
homogeneous solutions.
For two photons, if we make the following choice of homogeneous solutions,
$(B_1,C_1) = (2\k^2 E \theta (a) \d (u), 0)$ and
$(B_2,C_2) = (2\k^2 E \theta (a) \d (v), 0)$,
$h_{1\m\n}$ and $h_{2\m\n}$ will have support on the disjoint halfplanes,
$\{u=0,\, ({\rm sgn}\;a)y < 0\}$  and $\{v=0,\, ({\rm sgn}\;a) (y-a) > 0\}$,
allowing us to superpose the one-photon solutions to give the two-photon
solution:
\eqn\gravantipar{ ds^2 =   ds_0^2 -2\k^2 E\{  y ( \theta (y) -\theta (a))
\d (u)du^2
+ (a-y) (\theta (a-y) - \theta (a)) \d (v)dv^2\}.}

We now give an example in which the maximal number  (four) of
mutually non-parallel photons in TMG superpose, and construct
their exact solution.
Let two of the particles be antiparallel-moving in the $x$-direction with
energies $E_1$ and $E_2$, and the other two antiparallel
in the $y$-direction with
energies $E_3$ and $E_4$. Their trajectories and null halfplanes are given
respectively by
$(t=x,\,y=0),$ $\{t=x,\, y>0\};$ $(t=-x,\,y= a),$ $\{t= -x,\, y < a\};$
$(t=y,\,x= b ),$ $\{t=y ,\, x< b \};$
$(t= -y ,\, x=-c ),$ $\{t= -y ,\, x> -c \}.$
It is not difficult to check that if $a,\,b,\,c$ are all negative,
 the four halfplanes are disjoint,
and the four-photon solution can be obtained by superposing these four
one-photon solutions:
\eqn\fourphoton{\eqalign{ ds^2 = \,\,
& ds_0^2 + 2\k^2  \{ E_1 f(y) \d (t-x) {(dt-dx)}^2
+ E_2 f(-y+a) \d (t+x) {(dt+dx)}^2\cr
& + E_3 f(-x+b ) \d (t-y) {(dt-dy)}^2
+ E_4 f(x+c) \d (t+y) {(dt+dy)}^2 \}.  \cr}}

{\newsec{Geometrical  Approach.}}

In this section we obtain the Einstein gravity solutions of the previous
sections   by gluing together patches of three-dimensional Minkowski
space with Poincare transformations.  (In TMG,
since spacetime is not flat outside sources, this is not possible.)
Solutions in Einstein gravity were obtained this way
in {\djt} for timelike sources, and in {\ref\djtctc {S. Deser,
 R. Jackiw, and G. 't Hooft, {\it Phys. Rev. Lett.} {\bf 68} (1992) 267.}}
for tachyonic ones. Solutions for lightlike sources can be
obtained from the former by taking the  limit $\b \ra 1$ and
$m\ra 0$ with the energy fixed.
We recall that a static particle of
mass $m$ ($\le 2\pi$) is described by the
conical spacetime constructed by first  extracting a
wedge of angle $m$ (where we are setting $\k^2 =1$)
from the $x-y$ plane with vertex at the origin coinciding with the particle's
position. Points along the two edges  are then identified
by ${\bf x}^{\prime}  =  \Omega_m  {\bf x} $ where
$\Omega_m$ is a rotation matrix
\eqn\rotation{\Omega_m = \pmatrix{ 1 & 0  & 0\cr
                        0 & {\rm cos}\;m & {\rm sin}\;m  \cr
                       0  & - {\rm sin}\;m  & {\rm cos}\;m \cr} }
in the coordinate system  ${\bf x} = (t,x,y)$.
 To obtain the solution for  a particle
moving with speed $\b= {\rm tanh}\;\xi$  in the positive $x$ direction and
located at spacetime point ${\bf a}$, we boost and
translate the conical spacetime yielding the identification
\eqn\finitevid{{\bf x}^{\prime}  =  {\bf a} +(\L_{\xi} \Omega_m
\L_{\xi}^{-1}) ({\bf x}- {\bf a}  )  }
where $\L_{\xi}$ is the  boost,
\eqn\boostmatrix{\L_{\xi} = \pmatrix {{\rm cosh}\;\xi &  {\rm sinh}\;\xi & 0\cr
                            {\rm sinh}\;\xi     &  {\rm cosh}\;\xi &  0\cr
                     0     &  0  &1\cr }. }
For a photon we take the
infinite boost
limit with the energy $E =m\,{\rm  cosh}\;\xi$ fixed. With $ {\bf a}  ={\bf
0}$,
 {\finitevid} then becomes
\eqn \Ndef {{\bf x}^{\prime} = N_E\,\,  {\bf x},\;\;\;
 N_E  = \pmatrix { 1 +{1 \over 2}E^2 & -{1 \over 2} E^2  & E \cr
                          {1  \over 2}E^2  & 1- {1\over 2}E^2  & E  \cr
                            E & -E & 1  \cr},\;\;\; }
which in light cone coordinates corresponds to
\eqn\nullid{ (u^{\prime}, v^{\prime}, y^{\prime})  = (u, v  + 2E y +
E^2 u, y + E u).}
The matrices $N_E$ represent all Lorentz transformations which
leave the worldline
$u=y=0$ invariant; they therefore play the same role for a lightlike particle
as
the rotations $\O_m$ play in the case of a static
particle. If one now makes a cut along the
$u=0,\;y>0$ halfplane, and repastes with  the identification {\nullid},
one obtains the pure gravity limit of the
one-photon solution, {\gravid}, derived earlier.
We note that one can also
identify distinct halfplanes containing  the worldline
which are mapped into one another by $N_E$.
This leads to the
class of solutions known as null orbifolds which have been
discussed previously in the
context of string theory {\ref\hs{G. Horowitz and A. Steif, {\it Phys. Lett.}
{\bf B258} (1991) 91.}.

A system of particles can be described by composing the
respective one-particle identifications.  For two particles,
this yields
\eqn\twoid{\eqalign{{\bf x}^{\pr\pr} &= {\bf a_1} +
 \L_{1  }\O_{1} \L_{1}^{-1}
 ( {\bf x}^{\pr}- {\bf a_1})\cr
                               & = {\bf a_1} + \L_{1} \O_{1}
\L_{1}^{-1}
( ({\bf a_2}- {\bf a_1}) +
\L_{2} \O_{2} \L_{2}^{-1} ( {\bf x}-{\bf a_2}) ).\cr}}
This in turn is equivalent to the identification, {\finitevid},
 for a single composite particle
where $ T \equiv \L \O_m \L^{-1} =
\L_{1} \O_{1} \L_{1}^{-1}   \L_{2} \O_{2} \L_{2}^{-1} . $
$m$ is the total mass of the system, and $\L$ is some boost.
Taking the trace then yields
\eqn\mass{\eqalign{   {\rm cos}\;  m &
              = {1\over 2} ({\rm Tr}\;\O_m -1)
               = {1\over 2} ({\rm Tr}\;T -1)   \cr
 &  = {1\over 2} ({\rm Tr}\;   ( \L_{1} \O_{1} \L_{1}^{-1}
  \L_{2} \O_{2} \L_{2}^{-1})     -1).\cr } }
For two photons both moving   in the positive $x$-direction
 with energies
$E_1$ and $E_2$, this implies
$T = N_{E_1}N_{E_2} = N_{E_1+ E_2}. $ From {\mass},  we find as expected
that $m=0$. For
photons moving antiparallel in their center-of-momentum frame each with
energy $E$, the matrix becomes
$T = (\O_{\p}  N_E  \O_{-\p} ) N_E;$ {\mass} then implies
\eqn\antimass{ {\rm cos\;} m  =  1- 2E^2 + {1\over 2}E^4 .}
For small $E$, this reduces to $m=2E$,
the usual flat space mass addition formula.
 In order for the mass $m$ to be real (non-tachyonic),
the right hand side of {\antimass} must be in the
interval $[-1, 1].$ This in turn places an upper bound on the energy:
 $ E  <   E_{\rm max} = 2.$

As a last application, consider a  mixed system consisting of a photon
and a particle of finite mass $M$. Even though
we did not obtain the analytic form for the
metric  describing this system, we can still describe the
solution geometrically.
In  a frame in which the massive particle is at rest and the
photon is moving in the positive $x$ direction with energy $E$, we have
$T = N_E\O_M.$  Taking the trace, we find that the mass $m$
 of the composite system is given by
\eqn\mixedmass{  {\rm cos\;} m = {\rm cos\;} M - ({\rm sin\;} M)
E  +
{1\over 4}  (1-{\rm cos\;} M)  E^2.}
For small $M$ and $E$, this again coincides with the flat space formula,
$m^2 = M^2 + 2ME$. As before,  the condition that
the total mass be real places an upper bound on $E$
\eqn\mixedphysicality{ E <  E_{{\rm max}}  =  2 { {\rm sin\;} M +
\sqrt{ 2 (1-{\rm cos\;} M)} \over  1-{\rm cos\;} M}  .}
As expected we see that as $M\ra 0$, $E_{{\rm max}} \ra \infty.$
On the other hand, as $ M$ approaches its maximal value of $2\p$,
observe that $E_{{\rm max}}\ra 0$ (since ${\rm sin}\; M <0$),
 {\it i.e.,} no photon is permitted in this limiting cylindrical geometry.

{\newsec{Discussion.}

We have obtained a series of explicit solutions for both the general nonlinear
TMG model and its limiting cases of Einstein gravity and pure Chern-Simons
gravity in $D=3$, all generated by (at least) one  null matter source.  These
solutions have a number of possible applications.  First, as we will discuss
elsewhere {\dms},  the single photon metric can be used to
obtain the eikonal gravitational scattering amplitude for a timelike, massive
particle colliding at small angles with the rapid one, following the methods of
{\thooft}; because the plane waves are always impulsive, the metric can written
as a (singular) pure gauge, resulting in an Aharonov-Bohm type of
scattering.
Secondly, since our solutions involve moving particles and hence have nonzero
orbital angular momentum, they will permit us to analyze
{\ref\ctcus {S. Deser and A. Steif, ``Time Machines from Lightlike
Sources?,''
in preparation.}} the conditions under which closed timelike curves can
be present both in
Einstein gravity (as well as TMG) for quite different
sources than those used in previous
work {\djt, \djtctc, \ref\ctc{J.
Gott, {\it Phys. Rev. Lett.} {\bf 66} (1991) 1126;
A. Ori, {\it Phys. Rev. D} {\bf 44}
(1991) R2214; C. Cutler, {\it Phys.
Rev. D} {\bf 45} (1992) 487;   S. Carroll,  E. Farhi, and
A. Guth, {\it Phys. Rev. Lett.}
{\bf 68} (1992) 263, (E)3368; MIT preprint CTP-2117 (1992);
G. 't Hooft, {\it Class. Quantum Grav.}
{\bf 9} (1992)
1335; D. Kabat, MIT preprint CTP-2034 (1992).
}}.

\bigbreak\bigskip\bigskip\centerline{\bf Acknowledgements}\nobreak
We thank J. McCarthy for useful discussions in the early stages of this work.
S. D. thanks T. Dereli for
informative correspondence on the CSG problem.

This work was supported in part by NSF Grant No. PHY-88-0451.

\listrefs
\end